\documentclass{article}
\usepackage{t1enc}      

\usepackage{amsmath}
\usepackage{amssymb}
\usepackage{mathrsfs}  

\newcommand{\ua}{\underline a \,}
\newcommand{\ub}{\underline b \,}
\newcommand{\uc}{\underline c \,}

\newcommand{\uA}{\underline A \,}
\newcommand{\uB}{\underline B \,}
\newcommand{\uC}{\underline C \,}

\newtheorem{theorem}{Theorem}[section]
\numberwithin{equation}{section}

\begin{document}
\bibliographystyle{unsrt}

\title{On some global problems in the tetrad approach to quasi-local 
quantities}

\author{ L\'aszl\'o B Szabados \\
Research Institute for Particle and Nuclear Physics \\
H-1525 Budapest 114, P. O. Box 49, Hungary \\
e-mail: lbszab@rmki.kfki.hu }
\maketitle

\begin{abstract}
The potential global topological obstructions to the tetrad approach 
to finding the quasi-local conserved quantities, associated with 
closed, orientable spacelike 2-surfaces ${\cal S}$, are investigated. 
First we show that the Lorentz frame bundle is always globally 
trivializable over an open neighbourhood $U$ of any such ${\cal S}$ if 
an open neighbourhood of ${\cal S}$ is space and time orientable, and 
hence a globally trivializable $SL(2,{\mathbb C})$ spin frame bundle 
can also be introduced over $U$. Then it is shown that all the spin 
frames belonging to the same spinor structure on ${\cal S}$ have 
always the same homotopy class. On the other hand, on a 2-surface with 
genus $g$ there are $2^{2g}$ homotopically different Lorentz frame 
fields, and there is a natural one-to-one correspondence between these 
homotopy classes and the different $SL(2,{\mathbb C})$ spinor 
structures. 

\end{abstract}


\section{Introduction}
\label{sec-1}

It is known that, just because of the complete diffeomorphism 
invariance of the theory, finding the appropriate {\em notion} of 
energy-momentum and angular momentum of gravitating systems in general 
relativity is highly non-trivial. (For a recent review see e.g. 
\cite{Sz04}.) One possible approach to finding them is based on the 
tetrad formulation of GR. (Though the basic idea of using tetrad 
fields already appeared even in the early 1950s, the first who 
systematically investigated the conserved quantities in this formalism 
was probably M\o ller \cite{Mo61}. This formalism has a long history 
with extended bibliography, in which many of the classical results are 
rediscovered from time to time. However, instead of giving a complete 
bibliography of the field, we refer only to the reviews 
\cite{Go,Al:etal05}. Some of the huge number of papers using tetrad 
formalism and giving an extended list of references are 
\cite{Spa,Ne89a,Ne89b,Sz91,Sz92,And:etal,ObRu06a,It}.) 

In this approach the basic field variable is the (for the sake of 
simplicity, orthonormal) vector basis (or tetrad field) $\{E^a_{\ua}
\}$, ${\ua}=0,...,3$. The advantage of this approach is that although 
the standard expression for the contravariant form $\tau^{\alpha
\beta}$ of the canonical energy-momentum, derived from M\o ller's 
Lagrangian in some local coordinate system $\{x^\alpha\}$, is only 
{\em pseudotensorial}, the canonical spin $\sigma^{\mu\alpha\beta}$ 
and the canonical Noether current, built from $\tau^{\alpha\beta}$ 
and $\sigma^{\mu\alpha\beta}$ as 

\begin{equation}
C^\mu\bigl[K\bigr]:=\tau^{\mu\beta}K_\beta+\Bigl(\sigma^{\mu[
\alpha\beta]}+\sigma^{\alpha[\beta\mu]}+\sigma^{\beta[\alpha\mu]}
\Bigr)\partial_\alpha K_\beta \label{eq:1.1}
\end{equation}
with any vector field $K^a$, are independent of the actual coordinate 
system, i.e. they are {\em tensorial} \cite{Sz91,Sz92}. This current 
can be derived from a superpotential $\vee_e{}^{ab}$ through $\kappa 
C^a[K]-G^{ab}K_b=\frac{1}{2}\nabla_b(K^e\vee_e{}^{ab})$, which is also 
tensorial and is known as M\o ller's superpotential \cite{Mo61,Go}. 
(Here $\kappa:=8\pi G$ with Newton's gravitational constant $G$, and 
we use the conventions in which Einstein's equations take the form 
$G_{ab}=-\kappa T_{ab}$.) Moreover, the tensorial nature of $\vee_e
{}^{ab}$ makes it possible to introduce a {\em tensorial} 
energy-momentum too \cite{Sz91,Sz92}. Nevertheless, the final 
expressions still {\em do} depend on the actual choice of the tetrad 
field. 

However, there is another (and always overlooked) potential difficulty 
in this approach. Namely, the basic field variable is defined only on 
an open subset $U\subset M$ of the spacetime manifold, e.g. when $U$ 
is contractible, but in general such a $U$ cannot be extended to the 
whole $M$. Thus though the current $C^a[K]$ is a genuine vector field, 
it is defined only on the domain of the tetrad fields. The obstruction 
to the globality of such a tetrad field is the global non-triviality 
of the orthonormal frame bundle $B(M,O(1,3))$ over the spacetime 
manifold. Since in general this bundle is {\em not} trivial, the 
resulting expression for the gravitational energy-momentum density has 
the extra limitation that it is defined only on the local 
trivialization domains of the frame bundle. 

Instead of the orthonormal frames we could use normalized spin frames 
$\{{\cal E}^A_{\uA}\}$, ${\uA}=0,1$, too (see e.g. \cite{Spa,MaFr}). 
Let us consider the orthonormal frame field to be built from the spin 
frame, $E^a_{\ua}=\sigma^{{\uA}{\uA}'}_{\ua}{\cal E}^A_{\uA}\bar{\cal 
E}^{A'}_{{\uA}'}$, where $\sigma^{{\uA}{\uA}'}_{\ua}$ are the 
standard $SL(2,{\mathbb C})$ Pauli matrices divided by $\sqrt{2}$ 
(see e.g. \cite{PR}). Then, combining results of \cite{MaFr} and of 
\cite{Sz91,Sz92,Sz04}, for the dual of M\o ller's tensorial 
superpotential we obtain the following remarkable expression 

\begin{equation}
\frac{1}{4}\sigma^{\ua}_{{\uA}{\uB}'}E^e_{\ua}\vee_e{}^{ab}\frac{1}{2}
\varepsilon_{abcd}=u\bigl({\cal E}_{\uA},\bar{\cal E}_{{\uB}'}\bigr)
_{cd}+\overline{u\bigl({\cal E}_{\uB},\bar{\cal E}_{{\uA}'}\bigr)
_{cd}}, \label{eq:1.2}
\end{equation}
where overline denotes complex conjugation and $u(\lambda,\bar\mu)
_{cd}:=\frac{\rm i}{2}(\bar\mu_{C'}\nabla_{DD'}\lambda_C-\bar\mu_{D'}
\nabla_{CC'}\lambda_D)$ is known as the $SL(2,{\mathbb C})$ spinor 
form of the Nester--Witten 2-form. However, the use of spinors does 
not improve the frame dependence of the local energy-momentum 
expressions: the superpotential (\ref{eq:1.2}) still depends 
essentially on the actual spin frame. Indeed, as in the previous 
paragraph, it is defined only in the local trivialization domains of 
the spin frame bundle. Moreover, in the spinor approach, first we 
would have to ensure the existence of spinors (in the form of a 
spinor structure). 

This difficulty is more manifest if we want to associate the conserved 
quantities with {\em extended} domains of spacetime, i.e. when they are 
intended to be introduced globally or quasi-locally \cite{Sz04}. In 
particular, the tensorial superpotential for the canonical Noether 
current (\ref{eq:1.1}), i.e. the 2-form $K^e\frac{1}{4}\vee_e{}^{ab}
\frac{1}{2}\varepsilon_{abcd}$, could be integrated on a closed, 
orientable spacelike 2-surface ${\cal S}$ to obtain a potentially 
reasonable quasi-local energy-momentum and angular momentum expression 
{\em only if ${\cal S}$ is in a local trivialization domain of the 
frame bundle}. However, it is not a priori obvious that any such 
2-surface is in a trivialization domain, e.g. when ${\cal S}$ has 
large genus or surrounds some spacetime singularity (and, in 
particular, when it is not the boundary of some compact subset of $M$, 
and hence it is not contractible). If it were, then equation 
(\ref{eq:1.2}) would yield a {\em natural Lagrangian interpretation 
to all the quasi-local energy-momentum expressions that are based on 
the Nester--Witten 2-form}: they are charge integrals of Noether 
currents derived from M\o ller's tetrad Lagrangian. 

For the {\em global} trivializability of $B(M,O(1,3))$ over the whole 
spacetime manifold $M$ Geroch gave necessary and sufficient conditions 
\cite{Ge1}: $(M,g_{ab})$ must admit a spinor structure, which condition 
in a time and space orientable spacetime is known to be equivalent to 
the vanishing of the second Stiefel--Whitney class of $M$ 
\cite{Mi,Bi,PR}. (For a number of sufficient conditions of the 
existence of a spinor structure, e.g. the global hyperbolicity, see 
\cite{Ge2}.) However, from the point of view of quasi-local quantities 
it would be enough if we could ensure that the 2-surface ${\cal S}$ is 
in {\em some} local trivialization domain of the Lorentz or spin frame 
bundles, independently of the global topological properties of subsets 
of $M$ that are `far' from our 2-surface. Thus now we are interested 
only in the ``quasi-local trivializability`` of the frame bundle $B(M,
O(1,3))$. 

If ${\cal S}$ is in such a domain, then the next question is how to 
choose the global frame field. The frame should not be fixed, constant 
basis (or gauge) transformations must be allowed. A remarkable 
property of the integral of the superpotential 2-form $K^e\frac{1}{4}
\vee_e{}^{ab}\frac{1}{2}\varepsilon_{abcd}$ on ${\cal S}$ is that it 
depends {\em only on the value of the tetrad field on ${\cal S}$, but 
independent of the way in which it is extended off the 2-surface}. In 
particular, it is an algebraic expression of $K^a$ and the tetrad field 
and its first derivative {\em tangential} to ${\cal S}$. Thus we need 
a {\em gauge condition}, e.g. in the form of a system of linear partial 
differential equations {\em only on ${\cal S}$}, admitting a six 
parameter family of solutions, and yielding the rotational freedom of 
the constant orthonormal basis of Minkowski spacetime. Therefore, it 
is natural to expect that the gauge condition, as a differential 
equation, is built up {\em only from the intrinsic and extrinsic 
geometry of the 2-surface}. 

The general strategy to prove the existence of its solutions could be 
to search for the `preferred' frame in the form $E^a_{\ua}={}_0E^a
_{\ub}\Lambda^{\ub}{}_{\ua}$, where $\{{}_0E^a_{\ua}\}$ is a fixed 
frame field and $\Lambda:{\cal S}\rightarrow SO_0(1,3)$, a space and 
time orientation preserving Lorentz matrix valued function on ${\cal 
S}$. The calculations would be simpler if we could write $\Lambda
^{\ua}{}_{\ub}=\exp(\lambda)^{\ua}{}_{\ub}:=\delta^{\ua}_{\ub}+
\lambda^{\ua}{}_{\ub}+\frac{1}{2}\lambda^{\ua}{}_{\uc}\lambda^{\uc}
{}_{\ub}+ ...$ for some $so(1,3)$ Lie algebra valued function $\lambda
:{\cal S}\rightarrow so(1,3)$. However, this can be done precisely 
{\em when the homotopy class of the frame fields $\{{}_0E^a_{\ua}\}$ 
and $\{E^a_{\ua}\}$ is the same}, e.g. when every continuous map 
$\Lambda:{\cal S}\rightarrow SO_0(1,3)$ is homotopic to the identity 
transformation. (Here $SO_0(1,3)$ denotes the connected component of 
the Lorentz group $O(1,3)$.) 

If we use spinors then there is the additional difficulty that there 
might be different spinor structures. Then the `preferred' spin frame 
field can be written as ${\cal E}^A_{\uA}={}_0{\cal E}^A_{\uB}A^{\uB}
{}_{\uA}$ for some given spin frame field $\{{}_0{\cal E}^A_{\uA}\}$ 
and gauge transformation $A^{\uA}{}_{\uB}=\exp(a)^{\uA}{}_{\uB}:=
\delta^{\uA}_{\uB}+a^{\uA}{}_{\uB}+\frac{1}{2}a^{\uA}{}_{\uC}a^{\uC}
{}_{\uB}+...$ with some $a:{\cal S}\rightarrow sl(2,{\mathbb C})$ 
precisely when the spin frames $\{{}_0{\cal E}^A_{\uA}\}$ and 
$\{{\cal E}^A_{\uA}\}$ belong to the {\em same} spinor structure and 
have the same homotopy class. 
Thus we should know the different $SL(2,{\mathbb C})$ spinor 
structures and the homotopy classes of the global Lorentz and $SL(2,
{\mathbb C})$ spin frame fields (or, equivalently, of the Lorentz and 
spin gauge transformations) on ${\cal S}$. 

The aim of the present paper is to clarify the specific global problems 
raised in the previous paragraphs. In particular, we want to find the 
conditions under which the quasi-local conserved quantities can be 
introduced in the tetrad formalism of general relativity even without 
imposing all the usual {\em global} topological restrictions on the 
{\em whole} spacetime. We will show that an orientable closed spacelike 
2-surface ${\cal S}$ is always contained in a trivialization domain 
$U$ of the orthonormal frame bundle if at least an open neighbourhood 
of ${\cal S}$ is space and time orientable, and hence a trivializable 
spin frame bundle can also be introduced over $U$. Thus the `quasi-local 
trivializability' of the Lorentz frame bundle is ensured by the 
orientabilities: in addition to the orientability of the 2-surface and 
the time and space orientability of {\em some} open neighbourhood of 
the 2-surface {\em no} extra global topological restriction is needed 
(in contrast to the global trivializability of the frame bundles over 
the {\em whole} spacetime manifold). In particular, in the tetrad 
approach to general relativity the quasi-local conserved quantities 
may be associated with a given 2-surface even if the Lorentz frame 
bundle is not globally trivializable over the whole spacetime manifold 
(an example for such a space and time orientable spacetime is given in 
\cite{Pe}), or even if the spacetime is not globally space and time 
orientable. We clarify the homotopy classes of the globally defined 
Lorentz and $SL(2,{\mathbb C})$ gauge transformations on ${\cal S}$ 
too. We find that in a given spinor structure there is 
always one homotopy class of the spin gauge transformations. However, 
on surfaces with genus $g$, there are $2^{2g}$ different homotopy 
classes of the Lorentz gauge transformations, and these are in a 
natural one-to-one correspondence with the different $SL(2,{\mathbb 
C})$ spinor structures. Thus, in the construction of the quasi-local 
conserved quantities on topological 2-spheres in the tetrad formalism 
none of the obstructions above can occur, and both the `preferred' 
spin and Lorentz frames can always be searched for in the form ${\cal 
E}^A_{\uA}={}_0{\cal E}^A_{\uB}\exp(a)^{\uB}{}_{\uA}$ and $E^a_{\ua}=
{}_0E^a_{\ub}\exp(\lambda)^{\ub}{}_{\ua}$, respectively, even for 
any fixed ${}_0{\cal E}^A_{\uB}$ and ${}_0E^a_{\ub}$ and some $a:
{\cal S}\rightarrow sl(2,{\mathbb C})$ and $\lambda:{\cal S}
\rightarrow so(1,3)$. On the other hand, for higher genus ($g\geq1$) 
surfaces a choice for the spinor structure, or, equivalently, for 
the homotopy class of the Lorentz frame field, must also be made. 
This choice should be a part of the gauge condition. 

In the next section we recall some known facts about the global 
properties of closed orientable 2-surfaces ${\cal S}$ that we need in 
what follows, and briefly discuss the normal bundle of such spacelike 
2-surfaces in $M$. Then, in section 3, the global trivializability of 
the Lorentz and spin frame bundles over an open neighbourhood $U$ 
of arbitrary closed, orientable spacelike 2-surfaces is proven. 
Finally, in section 4, the homotopy classes of the global frame 
fields, or, equivalently, of the global spin and Lorentz basis 
transformations on these 2-surfaces are determined. 

Here we adopt the abstract index formalism of \cite{PR}, and only 
underlined indices take numerical values. Our basic differential 
geometric reference is \cite{KN}, and we use the terminology of 
\cite{Steen} in homotopy theory.


\section{Closed spacelike 2-surfaces}
\label{sec-2}

Let ${\cal S}$ be a closed, orientable two-dimensional smooth manifold, 
and let $g$ denote its genus. Since the geometry of the different 
connected components of ${\cal S}$ are independent of each other, for 
the sake of simplicity (and without loss of generality) we assume that 
${\cal S}$ is connected. 

If $g\geq1$, then let $\{a_i,b_i\}$, $i=1,...,g$, be a canonical 
homology basis on ${\cal S}$; i.e. they are closed, homotopically 
inequivalent noncontractible curves on ${\cal S}$. Then the 
fundamental group of ${\cal S}$ is $\pi_1({\cal S})=\langle a_i,
b_i\vert \,\,\prod^g_{i=1}a_i\cdot b_i\cdot a^{-1}_i\cdot b^{-1}_i$ 
$=1\,\rangle$; i.e. $\pi_1({\cal S})$ is generated by the $2g$ 
elements $a_i$, $b_i$ with the only relation that the product of 
all the commutants $a_i\cdot b_i\cdot a^{-1}_i\cdot b^{-1}_i$ is 
homotopically trivial. (Here $a\cdot b$ denotes the composition of 
the closed curves $a$ and $b$ in the sense of homotopy theory, and 
$1$ is the identity element of the fundamental group, being 
represented by a closed curve in ${\cal S}$ homotopic to a point.) 
Since the first homology group (with integer coefficients), $H_1
({\cal S})$, is the abelianization of $\pi_1({\cal S})$, the form 
of its general element in additive notation is $\sum_{i=1}^g(m^ia_i
+n^ib_i)$ for some $m^i,n^i\in{\mathbb Z}$. Then any group 
homomorphism $\phi:H_1({\cal S})\rightarrow{\mathbb Z}_2$ is 
characterized completely by the values $\phi(a_i)$ and $\phi(b_i)$. 
Therefore, the cohomology group $H^1({\cal S},{\mathbb Z}_2)={\rm 
Hom}(H_1({\cal S}),{\mathbb Z}_2)$, consisting of all the group 
homomorphisms of $H_1({\cal S})$ into ${\mathbb Z}_2$, has $2^{2g}$ 
elements \cite{AN}. 

Next suppose that ${\cal S}$ is embedded as a smooth spacelike 
submanifold in the Lorentzian spacetime manifold $M$. Let $t^a$ and 
$v^a$ be timelike and spacelike unit normals to ${\cal S}$, 
respectively, satisfying $t^av_a=0$. Obviously, these unit normals are 
not unique, since there is a gauge freedom $(t^a,v^a)\mapsto (t^a\cosh 
u+v^a\sinh u,t^a\sinh u+v^a\cosh u)$ for any function $u:{\cal S}
\rightarrow{\mathbb R}$. However, $\Pi^a_b:=\delta^a_b-t^at_b+v^av_b$ 
is well defined: it is the orthogonal projection to ${\cal S}$ (by 
means of which e.g. the induced metric is defined as $q_{ab}:=\Pi^c_a
\Pi^d_bg_{cd}$). 

\begin{theorem}
  If ${\cal S}$ is orientable and an open neighbourhood of ${\cal S}$ 
  in $M$ is time and space orientable (which we assume in what follows), 
  then the unit normals $t^a$ and $v^a$ can (and, in the present paper, 
  will) be chosen to be globally well defined on ${\cal S}$ with 
  future pointing $t^a$ and (whenever defined) outward pointing $v^a$. 
\label{th:2.1}
\end{theorem}

\noindent
{\it Proof}: 
Let $W\subset M$ be an open neighbourhood of ${\cal S}$ which is time 
and space orientable. Then the existence of a globally defined timelike 
normal $t^a$ follows directly from the time orientability of $W$. By 
the orientability of ${\cal S}$ there is a nowhere vanishing area 
2-form $\varepsilon_{ab}$ on ${\cal S}$, and from the time and space 
orientability of $W$ its orientability, and hence the existence of a 
nowhere vanishing volume 4-form $\varepsilon_{abcd}$ on $W$, follows. 
Then $v^a:=\pm\frac{1}{2}\varepsilon^a{}_{bcd}t^b\varepsilon^{cd}$, 
depending on the choice of the orientation, is the desired globally 
defined spacelike normal. \hfill $\Box$

\medskip

A simple consequence of this theorem is that an open neighbourhood 
$U\subset M$ of ${\cal S}$ can be foliated by smooth spacelike 
hypersurfaces $\Sigma_t$, $t\in(-\tau,\tau)$ for some $\tau>0$, such 
that the leaves of this foliation are homeomorphic to ${\cal S}\times
(-\epsilon,\epsilon)$, $\epsilon>0$, and ${\cal S}$ is embedded in 
$\Sigma_0$ as $({\cal S},0)$. Hence $U$ is homeomorphic to ${\cal S}
\times(-\epsilon,\epsilon)\times(-\tau,\tau)$, and thus its homotopy 
retract is ${\cal S}$. 

By Theorem 2.1 the normal bundle $N{\cal S}$ of ${\cal S}$ is globally 
trivializable, and this trivialization is provided by the globally 
defined normals $\{t^a,v^a\}$. ${\mathbb V}^a({\cal S})$, the 
restriction to ${\cal S}$ of the tangent bundle $TM$ of $M$, has a $g
_{ab}$-orthogonal global decomposition: it is the direct sum of the 
tangent and normal bundles of the 2-surface, ${\mathbb V}^a({\cal S})
=T{\cal S}\oplus N{\cal S}$. In the next section we will see that 
${\mathbb V}^a({\cal S})$ is also globally trivializable, but this 
does {\em not} imply that the tangent bundle $T{\cal S}$ is also 
globally trivializable unless ${\cal S}$ is a torus. 

The triple $({\mathbb V}^a({\cal S}),g_{ab},\Pi^a_b)$ will be called 
the Lorentzian vector bundle over ${\cal S}$. The bundle of frames 
adapted to ${\cal S}$, i.e. the set of the orthonormal frames $\{e^a
_{\ua}\}$, where $e^a_1$ and $e^a_2$ are tangent, while $e^a_0$ and 
$e^a_3$ are orthogonal to ${\cal S}$, is an $SO(2)\times SO(1,1)$ 
principal fibre bundle, and is {\em not} globally trivializable in 
general. Its double covering spin frame bundle is just the GHP frame 
bundle with the structure group $GL(1,{\mathbb C})$. For some of its 
global properties see \cite{JoSz}. 


\section{The triviality of the Lorentz and spin frame bundles over 
a neighbourhood of ${\cal S}$}
\label{sec-3}

By our assumption an open neighbourhood $W\subset M$ of ${\cal S}$ is 
space and time orientable, and hence the restriction $B(W,O(1,3))$ of 
the orthonormal frame bundle $B(M,O(1,3))$ to $W$ is reducible to its 
subbundle $B(W,SO_0(1,3))$. By the next proposition its restriction 
to an appropriate open neighbourhood $U\subset W$ of ${\cal S}$ is 
isomorphic to a product bundle: 

\begin{theorem}
  There exists an open neighbourhood $U\subset M$ of ${\cal S}$ such 
  that the principal fibre bundle $B(U,SO_0(1,3))$ is globally 
  trivializable over $U$. \label{th:3.1}
\end{theorem}

\noindent
{\it Proof}: 
By the first consequence of Theorem 2.1 ${\cal S}$ has an open 
neighbourhood $U$ which is foliated by smooth spacelike hypersurfaces 
$\Sigma_t$, and let $t^a$ denote their future pointing unit timelike 
normal. Since any orientable 3-manifold is parallelizable \cite{Pa} 
(see also \cite{Steen}), there exists a globally defined orthonormal 
triad field $\{E^a_1,E^a_2,E^a_3\}$ on $\Sigma_0$. Extending this 
triad field along the integral curves of the timelike normal vector 
field $t^a$ in some smooth way to the other leaves of the foliation, 
$\{t^a,E^a_1,E^a_2,E^a_3\}$ provides a global trivialization of $B(U,
SO_0(1,3))$ on $U$. \hfill $\Box$

\medskip

Therefore, on some open space and time orientable neighbourhood of 
every closed, orientable 2-surface ${\cal S}$ the canonical Noether 
current and the corresponding tensorial superpotential are always 
well defined. 

The trivializability of $B(U,SO_0(1,3))$ implies the existence of a 
globally trivializable principal fibre bundle $\tilde B(U,SL(2,
{\mathbb C}))$ and a base point preserving surjective 2--1 bundle map 
$E:\tilde B(U,SL(2,{\mathbb C}))\rightarrow B(U,SO_0(1,3))$ taking the 
right action of $SL(2,{\mathbb C})$ on $\tilde B(U,SL(2,{\mathbb C}))$ 
to the right action of $SO_0(1,3)$ on $B(U,SO_0(1,3))$. Thus $E$ 
defines a spinor structure on the tangent bundle $TU$ of $U$, 
considered to be a spacetime manifold on its own right. Nevertheless, 
in general there might be other, inequivalent spinor structures on 
$TU$, labeled by the elements of the cohomology group $H^1(U,{\mathbb 
Z}_2)=H^1({\cal S},{\mathbb Z}_2)$. Hence the number of the 
inequivalent spinor structures on $TU$ is $2^{2g}$. By the next 
statement, however, the corresponding spin frame bundles as abstract 
principle fibre bundles over $U$ are all isomorphic to the trivial one. 

\begin{theorem}
 Any $SL(2,{\mathbb C})$ principal fibre bundle over $U$ is globally 
 trivializable. \label{th:3.2}
\end{theorem}

\noindent
{\it Proof}: 
First we show that any $SU(2)$ principal bundle over $U$ is globally 
trivializable. Thus let $B(U,SU(2))$ be any such bundle, let $\Sigma
_t$ be a foliation of $U$ by smooth spacelike hypersurfaces, and let 
$\xi^a$ be a vector field on $U$ such that $\xi^a$ is nowhere tangent 
to the leaves $\Sigma_t$ and the corresponding local 1-parameter 
family $\phi_t$ of diffeomorphisms maps the leaves of the foliation 
to leaves. Then, using this $\phi_t$, every point of $U$ can be 
represented by a pair $(t,p)$, where $p\in\Sigma_0$. It is known (see 
problem 18 of lecture 4 in \cite{Mo}) that every $SU(2)$-bundle over 
an orientable 3-manifold is globally trivializable, and hence admits 
a global cross section. (For a proof of this trivializability, using 
the triangulability of $\Sigma_0$, the arcwise connectedness of $SU
(2)$ and that $\pi_1(SU(2))=0$ and $\pi_2(SU(2))=0$, see \S29 of 
\cite{Steen}.) Thus let $B(\Sigma_0,SU(2))$ be the restriction of $B(U,
SU(2))$ to $\Sigma_0$, and let $\sigma_0:\Sigma_0\rightarrow B(\Sigma
_0,SU(2))$ be a global cross section. Then $\sigma:U\rightarrow B(U,
SU(2)):$ $(t,p)\mapsto\sigma_0(p)$ is a global cross section of $B(U,
SU(2))$, i.e. the bundle $B(U,SU(2))$ is trivializable over $U$. 
However, this implies the global trivializability of any principal 
fibre bundle $B(U,SL(2,{\mathbb C}))$ too, because any global cross 
section of any of its reduced subbundle $B(U,SU(2))\subset B(U,SL(2,
{\mathbb C}))$ is a global cross section of $B(U,SL(2,{\mathbb C}))$ 
as well. \hfill $\Box$

\medskip

Another way of proving the trivializability of the spin frame bundle 
over $U$ could be based on Geroch's theorem \cite{Ge1}: by theorem 3.1 
$U$, as a spacetime manifold, admits a globally defined orthonormal 
tetrad field, and hence it admits a spinor structure in the form of a 
globally trivializable $SL(2,{\mathbb C})$-spin frame bundle. 

Let us fix a spinor structure $E$ on $TU$ and let ${\mathbb S}^A(U)$ 
be the vector bundle associated to $\tilde B(U,SL(2,{\mathbb C}))$ 
with the natural action of $SL(2,{\mathbb C})$ on ${\mathbb C}^2$. 
Let $\varepsilon_{AB}$ be the symplectic fibre metric thereon 
inherited through $E$. By the trivializability of $\tilde B(U,SL(2,
{\mathbb C}))$ the vector bundle of unprimed spinors, ${\mathbb S}^A
(U)$, is also globally trivializable. In general, however, the spinor 
structure $E$ does not necessarily coincide with the restriction to 
$U$ of the spacetime spinor structure (even if $M$ admits a spinor 
structure). 
By the global trivializability of ${\mathbb S}^A(U)$ (or rather of 
$\tilde B(U,SL(2,{\mathbb C}))$) there exist globally defined 
(normalized) spin frame fields $\{{\cal E}^A_{\uA}\}$ on $U$. The map 
$E$ links the global Lorentz and spin frames on $U$ in the standard 
way: $\sigma^{{\uA}{\uA}'}_{\ua}{\cal E}^A_{\uA}\bar{\cal E}^{A'}
_{{\uA}'}$ is identified with an orthonormal Lorentz frame field 
$E^a_{\ua}$ (as it was already done in the introduction in connection 
with equation (\ref{eq:1.2})). 


\section{The homotopy classes of the Lorentz and spin frame fields on 
${\cal S}$}
\label{sec-4}

Obviously, the theorems of the previous section imply the global 
trivializability of the pulled back principal bundles $B({\cal S},SO_0
(1,3))$, $\tilde B({\cal S},SL(2,{\mathbb C}))$ and of the vector 
bundles ${\mathbb V}^a({\cal S})$ and ${\mathbb S}^A({\cal S})$ to the 
2-surface ${\cal S}$, too; and the number of the inequivalent spinor 
structures on ${\mathbb V}^a({\cal S})$ is $2^{2g}$ (for the general 
notion of a spinor structure on a vector bundle, see e.g. \cite{LM}). 
However, it might be worth noting that since any closed orientable 
2-surface ${\cal S}$ can be triangulated and $SL(2,{\mathbb C})$ is 
arcwise connected and simple connected, any $SL(2,{\mathbb C})$ 
principal bundle over any such ${\cal S}$ is always globally 
trivializable (see \S29 of \cite{Steen}). 

In the present section we clarify the homotopy properties of the 
globally defined basis transformations $E^a_{\ua}\mapsto E^a_{\ua}
\Lambda^{\ua}{}_{\ub}$ and ${\cal E}^A_{\uA}\mapsto{\cal E}^A_{\uA}
A^{\uA}{}_{\uB}$ on the 2-surface, where $\Lambda:{\cal S}\rightarrow 
SO_0(1,3)$ and $A:{\cal S}\rightarrow SL(2,{\mathbb C})$, as maps 
from ${\cal S}$ into the groups in question, are smooth. If ${\cal 
S}$ is homeomorphic to $S^2$, then the homotopy classes of these 
transformations define just the second homotopy groups $\pi_2(SO_0
(1,3))$ and $\pi_2(SL(2,{\mathbb C}))$, respectively (see 
\cite{Steen}), which are well known to be trivial. Next we clarify 
these homotopy classes on 2-surfaces with $g\geq1$. 

\begin{theorem}
  Any smooth map $A:{\cal S}\rightarrow SL(2,{\mathbb C})$ is 
  homotopic to the identity map $I:p\mapsto{\rm diag}(1,1)$. 
\label{th:4.1}
\end{theorem}

\noindent
{\it Proof}:
First recall that the homotopy retract of $SL(2,{\mathbb C})$ is 
$S^3$, and hence we should determine the homotopy classes only of 
the smooth maps $A:{\cal S}\rightarrow S^3$. Since ${\cal S}$ is two 
dimensional and $A$ is smooth, it cannot be surjective. Thus there 
is a point $n\in S^3-A({\cal S})$, and let us introduce the standard 
polar coordinates $(r,\theta,\phi)$ on $S^3-\{n\}$ with the `north 
pole' at $n$. If in these coordinates the map $A$ is given by $p
\mapsto(r(p),\theta(p),\phi(p))$, then let us define the 1-parameter 
family of smooth maps $A_t:{\cal S}\rightarrow S^3$ by $A_t(p):=
(t\, r(p),\theta(p),\phi(p))$ for any $t\in[0,1]$. But this is a 
smooth homotopy between $A$ and the constant map taking all points 
of ${\cal S}$ into the `south pole' $(0,{\rm undetermined},{\rm 
undetermined})$ of $S^3$. \hfill $\Box$

\medskip
Therefore, any globally defined gauge transformation $A:{\cal S}
\rightarrow SL(2,{\mathbb C})$ is homotopic to the identity 
transformation, and hence any such transformation is globally 
generated by a Lie algebra valued function $a:{\cal S}\rightarrow 
sl(2,{\mathbb C})$ via $A^{\uA}{}_{\uB}=\exp(a)^{\uA}{}_{\uB}$. 
To clarify the homotopy properties of the Lorentz gauge 
transformations $\Lambda:{\cal S}\rightarrow SO_0(1,3)$ too, recall 
that topologically $SO_0(1,3)$ is $SO(3)\times{\mathbb R}^3\approx
{\mathbb R}P^3\times{\mathbb R}^3$, and hence its fundamental group 
is ${\mathbb Z}_2$. 

\begin{theorem}
  If the genus of ${\cal S}$ is $g$, then there are precisely 
  $2^{2g}$ homotopically different gauge transformations $\Lambda:
  {\cal S}\rightarrow SO_0(1,3)$. \label{th:4.2}
\end{theorem}

\noindent
{\it Proof}: 
Since the homotopy retract of $SO_0(1,3)$ is $SO(3)$, it is enough 
to prove the statement for pure rotations. Thus let $\Lambda:{\cal 
S}\rightarrow SO(3)$ be any given global gauge transformation. If 
$\gamma$ is any closed curve in ${\cal S}$, then $\Lambda\circ
\gamma$ is a closed curve in $SO(3)$, and let us define the index 
$i_\Lambda(\gamma)$ of $\gamma$ with respect to $\Lambda$ to be 1 
if $\Lambda\circ\gamma$ is homotopic to zero in $SO(3)$, and to be 
--1 otherwise. It is easy to see that $i_\Lambda(\gamma)=i_\Lambda
(\gamma')$ if $\gamma$ and $\gamma'$ are homotopic in ${\cal S}$. 
Moreover, since the homotopy class $[\Lambda\circ(\gamma\cdot
\gamma')]$ is just the product $[\Lambda\circ\gamma][\Lambda\circ
\gamma']$ in $\pi_1(SO(3))$, it follows that $i_\Lambda(\gamma
\cdot\gamma')=i_\Lambda(\gamma)i_\Lambda(\gamma')$ for any two 
closed curves $\gamma$ and $\gamma'$ with common endpoints. This 
implies that the index defines a group homomorphism $i_\Lambda:\pi
_1({\cal S})\rightarrow\pi_1(SO(3))\approx{\mathbb Z}_2$. Next we 
show that two gauge transformations, say $\Lambda$ and $\Lambda'$, 
are homotopic precisely when $i_\Lambda(\gamma)=i_{\Lambda'}
(\gamma)$ for any closed $\gamma$. 

To see this, suppose first that $\Lambda$ and $\Lambda'$ are 
homotopic; i.e. there is a 1-parameter family of gauge 
transformations $\Lambda_t:{\cal S}\rightarrow SO(3)$ such that 
${\cal S}\times[0,1]\rightarrow SO(3)$ $:(p,t)\mapsto\Lambda
_t(p)$ is continuous and $\Lambda(p)=\Lambda_0(p)$, $\Lambda'(p)=
\Lambda_1(p)$ for any $p\in{\cal S}$. Then for any $t\in[0,1]$ and 
closed curve $\gamma$ in ${\cal S}$ the map $\Lambda_t\circ\gamma:
S^1\rightarrow SO(3)$ defines a closed curve in $SO(3)$. Then, 
however, $\Lambda\circ\gamma$ and $\Lambda'\circ\gamma$ are 
homotopic in $SO(3)$ with the homotopy $\Lambda_t\circ\gamma$ 
between them. Therefore, $i_\Lambda(\gamma)=i_{\Lambda'}(\gamma)$. 

Conversely, let $\Lambda$ and $\Lambda'$ be global gauge 
transformations such that $i_\Lambda(\gamma)=i_{\Lambda'}(\gamma)$ 
for any closed curve $\gamma$. We will construct a homotopy between 
$\Lambda$ and $\Lambda'$. Let $p_0\in{\cal S}$ be a point where 
$\Lambda(p_0)\not=\Lambda'(p_0)$. (We may assume the existence of 
such a point, because otherwise the two transformations would be 
the same.) Let $\gamma:S^1\rightarrow{\cal S}$ $:s\mapsto\gamma(s)$ 
be an arbitrary closed curve with the starting and end point $p_0=
\gamma(0)=\gamma(1)$. Then since $i_\Lambda(\gamma)=i_{\Lambda'}
(\gamma)$, the closed curves $\Lambda\circ\gamma$, $\Lambda'\circ
\gamma: S^1\rightarrow SO(3)$ are homotopic, and hence there 
exists a continuous map $\Gamma:S^1\times[0,1]\rightarrow SO(3)$ 
such that $\Gamma(s,0)=\Lambda\circ\gamma(s)$ and $\Gamma(s,1)=
\Lambda'\circ\gamma(s)$ for any $s\in S^1$. In particular, this 
map defines the continuous curve $t\mapsto\Gamma(0,t)$ in $SO(3)$ 
from $\Lambda(p_0)$ to $\Lambda'(p_0)$. 
Then there exists a uniquely determined 1-parameter subgroup 
$t\mapsto R(p_0,t)$ of $SO(3)$ such that $R(p_0,0)={\rm Id}$, $R
(p_0,1)=(\Lambda(p_0))^{-1}\Lambda'(p_0)$ and the curve $\Lambda
_t(p_0):=\Lambda(p_0)R(p_0,t)$ between $\Lambda(p_0)$ and 
$\Lambda'(p_0)$ is homotopic to $t\mapsto\Gamma(0,t)$. (Here 
$(\Lambda(p_0))^{-1}$ is the inverse of the group element $\Lambda
(p_0)$ in $SO(3)$.) 
However, this $\Lambda_t(p_0)$ can be uniquely extended to a 
continuous family of curves $\Lambda_t(p)$, $p=\gamma(s)$, joining 
$\Lambda(p)=\Lambda_0(p)$ to $\Lambda'(p)=\Lambda_1(p)$. Note that 
this extension to all along $\gamma$ is globally possible just by 
the homotopy between $\Lambda\circ\gamma$ and $\Lambda'\circ\gamma$. 
In fact, $\Lambda_t(\gamma(s))$ is another homotopy (being 
equivalent to $\Gamma(s,t)$ above), but, apart from its overall 
orientation, it is completely determined by the two gauge 
transformations $\Lambda$ and $\Lambda'$. 
Finally, deforming the closed curve $\gamma$ throughout ${\cal S}$ 
and composing it with other curves $\gamma'$ we obtain a continuous 
map $\Lambda_t:{\cal S}\rightarrow SO(3)$, $t\in[0,1]$, which 
defines a homotopy between the gauge transformations $\Lambda$ and 
$\Lambda'$. 

Therefore, there is a natural one-to-one correspondence between 
the homotopy classes of the gauge transformations $\Lambda$ and the 
group homomorphisms $i_\Lambda:\pi_1({\cal S})\rightarrow\pi_1
(SO(3))\approx{\mathbb Z}_2$. If $g=0$, then, as we already saw, 
there is only one such homotopy class. Thus we may assume that 
$g\geq1$. To determine the number of the group homomorphisms $i
_\Lambda$, let us use the canonical homology basis $\{a_i,b_i\}$, 
$i=1,...,g$, of ${\cal S}$. Then $i_\Lambda$ is characterized 
completely by the values $i_\Lambda(a_i)$ and $i_\Lambda(b_i)$. 
Since $i_\Lambda$ is a homomorphism and ${\mathbb Z}_2$ is 
commutative, $i_\Lambda(a_i)i_\Lambda(b_i)i_\Lambda(a^{-1}_i)i
_\Lambda(b^{-1}_i)=1$ holds for all $i=1,...,g$, and hence $\prod
^g_{i=1}a_i\cdot b_i\cdot a^{-1}_i\cdot b^{-1}_i=1$ does not give 
any restriction on the values $i_\Lambda(a_i)$ and $i_\Lambda(b_i)$. 
Hence the number of the homomorphisms $\pi_1({\cal S})\rightarrow
{\mathbb Z}_2$, i.e. the number of the homotopy classes of the 
global Lorentz gauge transformations is $2^{2g}$. \hfill $\Box$ 

\medskip
Therefore, in contrast to the $SL(2,{\mathbb C})$ transformations, 
the general Lorentz transformations on a 2-surface with genus $g
\geq1$ cannot be written as $\exp(\lambda)^{\ua}{}_{\ub}$ for some 
$\lambda:{\cal S}\rightarrow so(1,3)$, because these are all 
homotopic to the identity transformation (the homotopy is $\exp(t\,
\lambda)$, $t\in[0,1]$). Consequently, a general Lorentz gauge 
transformation necessarily has the form $\exp(\lambda)^{\ua}{}
_{\ub}\Lambda^{\ub}_{(k)\uc}$, where $\Lambda^{\ua}_{(k)\ub}$, 
$k=1,...,2^{2g}$, are fixed gauge transformations with different 
homotopy classes. (Of course, one of them can be chosen to be the 
identity transformation.) 

Since the number of the different spinor structures and the number 
of the homotopy classes of the Lorentz gauge transformations is 
$2^{2g}$ for any $g$, one might conjecture that there is a deeper 
connection between the $SL(2,{\mathbb C})$ spinor structures and 
the homotopy classes of the Lorentz gauge transformations. The next 
theorem states that this expectation is correct.

\begin{theorem}
  There is a natural one-to-one correspondence between the spinor 
  structures on ${\mathbb V}^a({\cal S})$ and the homotopy classes 
  of the global orthonormal frame fields in ${\mathbb V}^a({\cal 
  S})$.  \label{th:4.3}
\end{theorem}

\noindent
{\it Proof}: 
Let $\{{\cal E}^A_{\uA}\}$ be a normalized spin frame in a given 
spinor structure, and let $A:{\cal S}\rightarrow SL(2,{\mathbb 
C})$ be any spin gauge transformation. By Theorem \ref{th:4.1} 
this is homotopic to the identity transformation, and hence there 
is a 1-parameter family of $SL(2,{\mathbb C})$ transformations, 
$A_t:{\cal S}\rightarrow SL(2,{\mathbb C})$, $t\in[0,1]$, such 
that $A_0(p)={\rm diag}\,(1,1)$ and $A_1(p)=A(p)$ at every $p\in
{\cal S}$. Then, however, the corresponding Lorentz gauge 
transformation, $\Lambda^{\ua}_t{}_{\ub}:=\sigma^{\ua}_{{\uA}
{\uA}'}A^{\uA}_t{}_{\uB}\bar{A}^{{\uA}'}_t{}_{{\uB}'}\sigma
^{{\uB}{\uB}'}_{\ub}$, is a homotopy between $\Lambda^{\ua}_1
{}_{\ub}$ and the identity transformation. 
Therefore, the spinor structure with any spin frame $\{{\cal E}^A
_{\uA}\}$ determines the homotopy class of the orthonormal Lorentz 
frame field $E^a_{\ua}:=\sigma^{{\uA}{\uA}'}_{\ua}{\cal E}^A_{\uA}
\bar{\cal E}^{A'}_{{\uA}'}$. 

Conversely, suppose that the Lorentz frames determined by the 
normalized spin frames $\{{\cal E}^A_{\uA}\}$ and $\{\tilde{\cal 
E}^A_{\uA}\}$ are homotopic. Then, however, we can always find a 
gauge transformation $A:{\cal S}\rightarrow SL(2,{\mathbb C})$ 
such that $\{{\cal E}^A_{\uA}\}$ and $\{\tilde{\cal E}^A_{\uA}A
^{\uA}{}_{\uB}\}$ determine the same Lorentz frame. Thus, without 
loss of generality, we can assume that $\{{\cal E}^A_{\uA}\}$ and 
$\{\tilde{\cal E}^A_{\uA}\}$ define the {\em same} Lorentz frame, 
and hence $\tilde{\cal E}^A_{\uA}=\pm{\cal E}^A_{\uA}$. However, 
${\cal E}^A_{\uA}$ and $-{\cal E}^A_{\uA}$ are homotopic spin 
frames, and hence they belong to the same spinor structure. 
Therefore, the correspondence between the spinor structures and 
the homotopy classes of the Lorentz frames is indeed one-to-one. 
\hfill $\Box$

\medskip
In particular, there are four homotopically different global frame 
fields and four different spinor structures on the Lorentzian vector 
bundle ${\mathbb V}^a({\cal S})$ over a torus ${\cal S}\approx S^1
\times S^1$. To see them, let ${\cal S}$ be a standard torus of radii 
$R$ and $r$, given explicitly by $t=0$, $x=(R+r\cos\phi)\cos\Phi$, 
$y=(R+r\cos\phi)\sin\Phi$ and $z=r\sin\phi$, $R>r>0$, in the 
Cartesian coordinates of the Minkowski spacetime. Then one frame 
field can be the restriction to ${\cal S}$ of the constant 
Cartesian frame field $\{E^a_{\ua}\}:=\{(\frac{\partial}{\partial 
t})^a,...,(\frac{\partial}{\partial z})^a\}$. 
The vectors of the second may be $e^a_0:=E^a_0$, $e^a_1:=\frac{1}
{R+r\cos\phi}(\frac{\partial}{\partial\Phi})^a$, $e^a_2:=\frac{1}
{r}(\frac{\partial}{\partial\phi})^a$ and $e^a_3:=\varepsilon{}
^a{}_{bcd}e^b_0e^c_1e^d_2$. Then the spatial basis vectors of 
these frames are connected with each other by 

\begin{eqnarray}
e^a_1\!\!\!\!&=\!\!\!\!&-\sin(\Phi) E^a_1+\cos(\Phi) E^a_2, 
 \nonumber \\
e^a_2\!\!\!\!&=\!\!\!\!&-\sin(\phi)\cos(\Phi) E^a_1-\sin(\phi)
 \sin(\Phi) E^a_2+\cos(\phi) E^a_3, \label{eq:4.1}\\
e^a_3\!\!\!\!&=\!\!\!\!&\cos(\phi)\cos(\Phi) E^a_1+\cos(\phi)
 \sin(\Phi) E^a_2+\sin(\phi) E^a_3. \nonumber
\end{eqnarray}
For the third and the fourth frames we choose $\tilde e^a_{\ua}:=
(e^a_0,\tilde e^a_1,\tilde e^a_2,e^a_3)$ and $\tilde{\tilde e}{}
^a_{\ua}:=(e^a_0,\tilde{\tilde e}{}^a_1,\tilde{\tilde e}{}^a_2,
e^a_3)$, respectively, where 

\begin{eqnarray}
\tilde e^a_1\!\!\!\!&:=\!\!\!\!&\cos(\Phi)e^a_1+\sin(\Phi)e^a_2, 
 \hskip 20pt 
\tilde e^a_2:=-\sin(\Phi)e^a_1+\cos(\Phi)e^a_2; \label{eq:4.2a} \\
\tilde{\tilde e}{}^a_1\!\!\!\!&:=\!\!\!\!&\cos(\phi)e^a_1+
 \sin(\phi)e^a_2, \hskip 20pt
\tilde{\tilde e}{}^a_2:=-\sin(\phi)e^a_1+\cos(\phi)e^a_2. 
 \label{eq:4.2b}
\end{eqnarray}
In the Cartesian coordinates for the canonical homology basis we 
choose the curves $a(\Phi):=(0,R\cos(\Phi),R\sin(\Phi),r)$ and $b
(\phi):=(0,R+r\cos(\phi),0,r\sin(\phi))$. Then by (\ref{eq:4.1}) 
the basis $\{e^a_{\ua}\}$ undergoes a complete $2\pi$ rotation with 
respect to $\{E^a_{\ua}\}$ in the 2-planes spanned by $E^a_1$ and 
$E^a_2$, and in the 2-planes spanned by $E^a_1$ and $E^a_3$ along 
the curves $a$ and $b$, respectively. Similarly, by (\ref{eq:4.2a}) 
the frame $\{\tilde e^a_{\ua}\}$ undergoes a $2\pi$ rotation with 
respect to $\{e^a_{\ua}\}$ along $a$, but remains unrotated along 
$b$; while $\{\tilde{\tilde e}{}^a_{\ua}\}$ is rotated with respect 
to $\{e^a_{\ua}\}$ along $b$, but remains unrotated along $a$. Thus, 
denoting the Lorentz matrices corresponding to (\ref{eq:4.1}), 
(\ref{eq:4.2a}) and  (\ref{eq:4.2b}), respectively, by $\Lambda$, 
$\tilde\Lambda$ and $\tilde{\tilde\Lambda}$, for the index of the 
closed curves $a$ and $b$ we obtain that $i_\Lambda(a)=i_\Lambda
(b)=-1$, $i_{\tilde\Lambda}(a)=-i_{\tilde\Lambda}(b)=-1$ and $i
_{\tilde{\tilde\Lambda}}(a)=-i_{\tilde{\tilde\Lambda}}(b)=1$, 
indicating that no two of the four Lorentz frames above are 
homotopic. 

It is easy to see that the spin frames $\{{\cal E}^A_{\uA}\}$, 
$\{\varepsilon^A_{\uA}\}$, $\{\tilde\varepsilon^A_{\uA}\}$ and 
$\{\tilde{\tilde\varepsilon}^A_{\uA}\}$ corresponding to the 
Lorentz frames $\{E^a_{\ua}\}$, $\{e^a_{\ua}\}$, $\{\tilde e^a
_{\ua}\}$ and $\{\tilde{\tilde e}{}^a_{\ua}\}$, respectively, 
belong to different spinor structures. For example, if, on the 
contrary, we assume that $\{\varepsilon^A_{\uA}\}$ and $\{\tilde
\varepsilon{}^A_{\uA}\}$ belong to the same spinor structure, then 
these spin frames would have to be connected by a globally defined 
$SL(2,{\mathbb C})$ transformation $\tilde A$. Apart from an overall 
sign, this would be fixed by (\ref{eq:4.2a}), and would be given by 
${\rm diag}(\exp(\frac{\rm i}{2}\Phi),\exp(-\frac{\rm i}{2}\Phi))$, 
yielding at the common starting and end point $a(0)=a(2\pi)$ of the 
closed curve $a$ that $\tilde\varepsilon^A_{\uA}(a(0))=-\tilde
\varepsilon^A_{\uA}(a(2\pi))$. 

\bigskip
The author is grateful to the referees for their valuable remarks 
and useful criticism. This work was partially supported by the 
Hungarian Scientific Research Fund (OTKA) grants T042531 and K67790.


\end{document}